\documentclass[11pt,a4paper]{article}
\usepackage[cal=boondox]{mathalfa}
\usepackage{comment}
\usepackage{jheppub}
\usepackage{mathtools}
\usepackage{float}
\usepackage{tikz-cd}

\date{}

\newcommand{\z}{\left}
\newcommand{\y}{\right}
\newcommand{\p}{\partial}

\newcommand{\aaa}{\alpha}

\newcommand{\bp}{\bar{\partial}}

 \usepackage[normalem]{ulem}
\usepackage[textsize=scriptsize,textwidth=2.5cm]{todonotes}

\newcommand{\sidenote}[1]{\textcolor{purple}{#1}}

\let\exporig\exp
\usepackage{physics}
\usepackage{enumitem}
\let\exp\exporig

\affiliation[\ensuremath{\gamma}]{Yau Mathematical Sciences Center, Tsinghua University, Beijing 100084, China}
\affiliation[\ensuremath{\tau}]{Department of Mathematical Sciences, Tsinghua University, Beijing 100084, China}

\usepackage{xspace}

\title{Asymptotic symmetries from  the string worldsheet}
\author{Zhengyuan Du$^{\gamma,\tau}$,}
\author{Kangning Liu$^{\gamma,\tau}$,}
\author{Wei Song$^{\gamma,\tau}$}

\abstract{
In IIB string theory on AdS$_3$ background with NS-NS fluxes, we show that Brown-Henneaux asymptotic Killing vectors can be derived by requiring both the worldsheet equations of motion and Virasoro constraints are preserved near the asymptotic boundary of the target spacetime. The charges on the worldsheet that generate the corresponding transformations can be written down in both the Lagrangian formalism and Hamiltonian formalism. This provides a method of studying asymptotic symmetry of the target spacetime directly from worldsheet string theories, without using results from the supergravity limit. As an example, we apply this method to flat spacetime in three dimensions and obtain the BMS$_3$ generators on the worldsheet theory. 

}

\begin{document}
\maketitle
\setlength{\parskip}{.3\baselineskip}
\section{Introduction}
Asymptotic symmetry plays an essential role in bottom-up holography. Assuming the existence of holographic duality,  the asymptotic symmetry of the gravitational theory in the bulk under certain boundary conditions should agree with the global symmetry of the putative dual quantum field theory in lower dimensions. 
In the seminal paper of \cite{Brown:1986nw}, the asymptotic symmetry of Einstein gravity with a negative cosmological constant in three dimensions is shown to be generated by two copies of Virasoro algebras, which indicates that the phase space of the gravity theory is the same as that of conformal field theory in two dimensions.  In addition, the Bekenstein-Hawking entropy of BTZ black holes \cite{Banados:1992wn} can be reproduced microscopically from the Cardy formula of a CFT$_2$ \cite{Strominger:1997eq}. 
These observations develop an approach to studying AdS$_3$/CFT$_2$ without having to know the underlying UV complete version of quantum gravity.
The strategy of studying holography from asymptotic symmetry has been generalized to spacetimes that are more closely related to the real world, including de Sitter spacetime \cite{Strominger:2001pn} and Kerr black holes \cite{Guica:2008mu}, etc. More recently, the asymptotic symmetry for flat spacetime, BMS symmetry \cite{bondi1962gravitational, sachs1962gravitational,Barnich:2006av},  has attracted increasing attention and led to the development of celestial holography \cite{Strominger:2013jfa,Strominger:2014pwa,Strominger:2017zoo}.

A precise realization of holography, however, requires a top-down approach, in which string theory is usually the starting point. So far, most examples of top-down holography deal with string theory on asymptotically AdS spacetimes. The development of holography for other backgrounds, though very important, is far behind\footnote{See \cite{Costello:2022jpg,Giveon:2017nie,Apolo:2019zai}, \cite{Chakraborty:2018vja, Apolo:2018qpq} for progresses on constructing flat holography, and Kerr/CFT in string theory respectively.}. Inspired by the role asymptotic symmetry played in the bottom-up holography,  
one natural and interesting question is how asymptotic symmetries appear in string theory. The answer to this question will help us understand the symmetry structure of the dual quantum field theory, provided that the latter exists, and thus make progress in the search for top-down holography beyond the AdS/CFT correspondence.

In this paper, we propose a method of constructing the asymptotic symmetry in the target spacetime directly from string worldsheet theories. For this purpose,  we revisit the WZW model \cite{Maldacena:2000hw,Maldacena:2000kv,Maldacena:2001km} which describes the worldsheet string theory of II B string theory on AdS$_3\times \mathcal N$ with NS-NS fluxes. It has been conjectured that superstring theory with $k=1$ unit of NS-5 brane charge is holographically dual to symmetric product theories \cite{Eberhardt:2019ywk}. For a generic value of $k$, a further deformation of the symmetric orbifold is required \cite{Eberhardt:2019qcl, Eberhardt:2021vsx, Balthazar_2022, martinec2021ads3s}. In this paper, we focus on string theory and will assume general $k$. 
Based on the asymptotic Killing vectors found by Brown-Hennaux \cite{Brown:1986nw},  
vertex operators corresponding to the Brown-Henneaux modes have been constructed previously in \cite{Giveon:1998ns,deBoer:1998gyt}. In this paper, we will first discuss how to find the asymptotic Killing vectors directly from the worldsheet theory without the knowledge of Brown-Henneaux.  Then we will further show how to systematically write down the corresponding operators that generate these transformations. 

Starting from the string worldsheet theory on a fixed background, we propose that a transformation of asymptotic symmetry in the target spacetime  has to satisfy the following two conditions: 
\begin{itemize}
\item The worldsheet equations of motion have to be preserved asymptotically, i.e. they have to vanish up to some specified order if expanded in terms of the radial coordinate.
\item The Virasoro constraints have to be satisfied asymptotically.
\end{itemize}
In the example of string theory on AdS$_3$, we will show that the above two conditions can be used to determine the asymptotic Killing vectors, which are indeed the Brown-Henneaux Virasoro generators. 
We use both the Lagrangian formalism and the Hamiltonian formalism to construct the worldsheet charges.  The results obtained in both formalisms are compatible with each other and with \cite{Giveon:1998ns,deBoer:1998gyt}.
We then apply this method to study the worldsheet theory of three-dimensional flat spacetime near null infinity. The above requirements allow us to reproduce the BMS$_3$ generators \cite{Barnich:2010eb}, and furthermore write down the charges on the worldsheet theory.

The layout of this paper is as follows: In section 2 we provide the basic setup of string theory on 
AdS$_3\times \mathcal N$ with NS-NS fluxes. In section 3, we study asymptotic symmetry in the supergravity limit. Under a certain choice of boundary conditions,  
we find that the Brown-Henneaux generators are accompanied by a gauge transformation, a fact that has not been fully spelled out in the literature.  In section 4, we interpret spacetime boundary conditions and asymptotic symmetry from the WZW model, and propose a strategy of finding the asymptotic symmetries from the worldsheet. We also discuss the Noether charges and Ward identity in the Lagrangian formalism. In section 5, we use the Hamiltonian formalism to rederive the symmetry generators, and the result is found to be consistent with previous results. In section 6, we derive the BMS$_3$ generators in three-dimensional flat spacetime from the string worldsheet.

\section{IIB string theory on AdS$_3\times \mathcal{N}$}
In this section, we review II B string theory on AdS$_3\times \mathcal N$ with NS-NS fluxes, with the purpose of setting up conventions. 
This theory has a weakly coupled worldsheet description, the three-dimensional part of which is the $SL(2,\mathbb{R})$ WZW model, which has been studied extensively in the literature. 
See \cite{Maldacena:2000hw,Maldacena:2000kv,Maldacena:2001km} for the spectrum,  \cite{Eberhardt:2019ywk} for the holographic duality at $k=1$, and \cite{Eberhardt:2019niq,Eberhardt:2021vsx} for generic $k$.  

For the purpose of asymptotic symmetry, it is convenient to consider the cylindrical boundary 
 so that the phase space of gravity contains the global AdS$_3$ and the  BTZ black holes. 
Under the holographic dictionary, global AdS$_3$ is identified as the NS-NS vacuum of the dual CFT$_2$, whereas massless BTZ is dual to the RR vacuum. More explicitly, the zero temperature BTZ background with $k$ NS-5 brane charge and $N$ NS-1 charge can be written as
\begin{equation}\label{1}
    \begin{aligned}
        d{s}^2&=\ell^2\left\{d \phi^2+\exp(2 \phi)\,d{u}\,d{v}\right\},\\
        {B}_{\mu\nu}&=-\frac{\ell^2}{2}\exp(2 \phi)\,d{u}\wedge d{v},\\
        e^{2 \phi}&=\frac{k}{N} e^{-2\phi_0}, \quad k=\ell^2/\alpha',
    \end{aligned}
\end{equation}
where we have omitted the internal spacetime. The lightcone coordinates ${u}\equiv{\varphi}+{t}$ and $v\equiv{\varphi}-{t}$
have the identification \begin{equation}\label{uvid}
    ({u},{v})\sim ({u}+2\pi,{v}+2\pi),
\end{equation}
so that the conformal boundary is a cylinder. 
The magnetic charge $k=\ell^2/\alpha'$ specifies how large the
 curvature scale is compared to the string scale. A small value of $k$ indicates strong stringy effects. 

The background \eqref{1} allows weakly coupled string worldsheet description. Using the plane coordinate on the worldsheet with $z\equiv\exp(i(\sigma-i\tau))$ and $\bar{z}\equiv\exp(-i(\sigma+i\tau))$,
the classical string worldsheet theory on \eqref{1} can be written in the conformal gauge as 
\begin{equation}\label{Action:AdS}
  S=\frac{1}{2\pi \aaa'}\int dz^2 {M}_{\mu\nu} \p  { X}^\mu \bar \p { X}^\nu =\frac{k}{ {2\pi}}\int d^2z\left\{\partial  \phi\bar{\partial} \phi+\exp(2 \phi)\bar{\partial}{u}\partial{v}\right\},
\end{equation}
where $d^2 z = dz\,d\bar{z}$, and we have
defined the combination
\begin{equation}
{M}_{\mu\nu} = { g}_{\mu\nu} +  B_{\mu\nu} \label{Mdef}. 
\end{equation}

At the quantum level, the level of the WZW model acquires a shift and the action reads  \cite{Giveon:1998ns,DiFrancesco:1997nk,Gerasimov1990WessZuminoWittenMA}
\begin{equation}\label{quantumaction}
    {S}=\frac{1}{2\pi}\int dz^2\z\{(k-2)\partial \phi\bar{\partial}\phi+k\exp(2 \phi)\bar{\partial}u\partial {v}-\frac{1}{4} \phi R_{ws}\y\},
\end{equation}
where $R_{ws}$ is the worldsheet curvature which vanishes on a flat worldsheet metric. The stress tensor is 
\begin{equation}\label{stressAdS}
\begin{aligned}
        &T_{ws}=-(k-2)\partial\phi\partial\phi-k\exp(2\phi)\partial u\partial v-\partial^2\phi,\\
        &\bar{T}_{ws}=-(k-2)\bar{\partial}\phi\bar{\partial}\phi-k\exp(2\phi)\bar{\partial} u\bar{\partial} v-\bar{\partial}^2\phi,
\end{aligned}
\end{equation}
with a central charge 
\begin{equation}
    c_{ws}=\bar{c}_{ws}=\frac{3k}{k-2}. 
\end{equation}
The worldsheet Virasoro generators are the Fourier mode of the worldsheet stress tensor  \begin{equation} L_n= {\frac{1}{2\pi i}}\oint dz z^{n+1} T_{ws},\quad \bar{L}_n= {-\frac{1}{2\pi i}}\oint d\bar{z}\bar{z}^{n+1} \bar{T}_{ws}. \end{equation}

Spacetime isometries correspond to global symmetries on the worldsheet. 
In particular, the worldsheet Noether currents of translational symmetries along the target-space coordinates ${u}$ and ${u}$ are given by
\begin{equation}\label{tildej}
    {j}_0=k\exp(2 \phi)\partial{v},\quad 
    \bar{j}_0=k\exp(2 \phi)\bar{\partial}{u}.
\end{equation}
The Noether charge is given by the zero mode of the currents
\begin{equation}
    J_0=  -{\frac{1}{2\pi}}  \oint dz {j}_0 {(z)},\quad {\bar{J}}_0=  {\frac{1}{2\pi}}  \oint d\bar{z}\bar{j}_0 (\bar z) .
\end{equation}
Using the Wakimoto variables \footnote{Note that the worldsheet fields in this paper is not rescaled as in the literature. A quick way to fix the coefficient of the dilaton term in \eqref{quantumaction} and \eqref{betagamma} is to require that the interaction term $\beta \bar\beta e^{-2\phi}$ to have conformal weight $(1,1)$ under the resulting stress tensor. We thank L. Apolo for discussions regarding this point. }, the worldsheet theory \eqref{Action:AdS} can equivalently be written as
\begin{equation}\label{betagamma}
    S^\prime=\frac{1}{ {2\pi}}\int d^2z\left\{(k-2)\partial \phi\bar{\partial} \phi+k\beta\bar{\partial}{u}+k\bar{\beta}\partial {v}-k\exp(-2 \phi)\beta\bar{\beta}-\frac{1}{4}  \phi R_{ws} \right\}.
\end{equation}
After using the on-shell relation $k\beta= {{j}}_0$ and $k\bar{\beta}= {\bar{j}}_0$, the action \eqref{betagamma} goes back to \eqref{Action:AdS}. In the large $\phi$ limit, the interaction term can be ignored and the above theory \eqref{betagamma} consists of a free boson $\phi$, a chiral $(\beta, u)$ system, and an anti-chiral $(\bar\beta, v)$ system, with the free field OPEs, 
 \begin{equation}\label{opeAdS}
    \begin{aligned}
                       & \phi(z,\bar{z}) \phi(w,\bar{w})\sim-\frac{1}{2(k-2)}\log|z-w|^2,\\ 
                       &{j}_0(z){u}(w)\sim -\frac{1}{z-w},\quad{\bar{j}}_0(\bar{z}){v}(\bar{w})\sim -\frac{1}{\bar{z}-\bar{w}}.
    \end{aligned}
\end{equation}
The last line is also the 
Ward identity for the currents ${j}_0,\,{\bar j}_0$.

\section{Asymptotic symmetry from supergravity}

The study of boundary conditions and asymptotic symmetries of Einstein gravity in three dimensions with negative cosmological constant has been very fruitful. Brown and Henneaux \cite{Brown:1986nw} imposed Dirichlet boundary conditions and found that the asymptotic symmetry group is the same as a two-dimensional conformal field theory, and hence can be regarded as a precursor of the  AdS$_3$/CFT$_2$ correspondence. It was found in \cite{Porfyriadis:2010vg} that a slightly weaker version of Dirichlet boundary conditions can allow more choices of the asymptotic Killing vectors \cite{Guica:2008mu}, but still lead to the same asymptotic symmetry group.
Under the Dirichlet-Neumann boundary conditions \cite{Compere:2013bya}, the asymptotic symmetry was found to be that of the warped conformal field theory in two dimensions. See \cite{Troessaert:2013fma,Grumiller_2016} for more discussions of the boundary conditions for pure Einstein gravity.

In this section, we consider the boundary conditions for the IIB supergravity, with the $B$-field taken into account as well. As shown below, we find consistent boundary conditions under which asymptotic Killing vectors have to be accompanied by non-vanishing gauge transformations at the boundary. The gauge transformations will also make sure that the Virasoro generators can be realized as chiral currents on the string worldsheet. The crucial role of the gauge transformation is very similar to the story of Einstein-Maxwell gravity in two dimensions \cite{Hartman:2008dq}, and the CSS boundary conditions \cite{Compere:2013bya}.

After dimensional reduction to three dimensions, the bosonic part of the action of IIB supergravity in the Einstein frame  can be written as
\begin{equation}\label{sugraaction}
	S ={1\over 16\pi G}\int d^{3}x \sqrt{-g_E} \z(R-4\p_\mu \Phi\p_\nu \Phi-  {e^{-8\Phi}\over 12} H_{\mu\nu\lambda}H^{\mu\nu\lambda}+\frac{4}{\ell^2}e^{4\Phi}\y) ,
\end{equation}
where $H=dH$ is the field strength of the NS-NS field $B_{\mu\nu}$, $\Phi $ is the dilation, and the metric in the Einstein frame $g_E$ is related to that in string frame by $g_E=e^{-4\Phi}g$. Classical solutions with cylindrical boundary include global AdS$_3$ spacetime and the BTZ blackholes. More generally, we can find a generalization of Ba\~nados metric  
\begin{equation}\label{BTZ}
    \begin{aligned}
        &ds^2=\ell^2\Big\{d\phi^2+(\exp(2\phi)+\exp(-2\phi)\mathcal{L}(u)\bar{\mathcal{L}}(v))dudv+\mathcal{L}(u)du^2+\bar{\mathcal{L}}(v)dv^2\Big\},\\
        &B_{\mu\nu}=-\frac{\ell^2}{2}\Big\{\exp(2\phi)+\exp(-2\phi)\mathcal{L}(u)\bar{\mathcal{L}}(v)\Big\}du\wedge dv,\\
        &e^{2 \phi}=\frac{k}{N} e^{-2\phi_0},\quad (u,v)\sim(u+2\pi,\, v+2\pi),
    \end{aligned}
\end{equation}
where $\mathcal{L}=\bar{\mathcal{L}}=-{1\over4}$ corresponds to global AdS$_3$ and constant $\mathcal{L}\ge0,\,\bar{\mathcal{L}}\ge0$ corresponds to BTZ black holes.
Boundary conditions for the supergravity theory on asymptotic AdS$_3\times \mathcal{N}$ background can be conveniently written in the Fefferman-Graham gauge, 
\begin{equation}\label{FG}
\begin{aligned}
    ds^2&=d\phi^2 +e^{2\phi}  \Big( g^{(0)}_{ij}(x^i)+e^{-2\phi }g^{(2)}_{ij}(x^i) +\cdots  \Big) dx^i dx^j  ,\quad i,j\in \{u,v\},\\
     B&=e^{2\phi} \Big( b^{(0)}_{ij}(x^i)+e^{-2\phi }b^{(2)}_{ij}(x^i) +\cdots \Big) dx^i\wedge dx^j,
\end{aligned}  
\end{equation}
where we have omitted the extra dimensions that do not play any role.
To discuss the boundary conditions, we note that there is a cancellation in the $uv$ component between the metric and $B-$field i.e. $g_{uv}+B_{uv}=0$, for all solutions described by \eqref{BTZ}. 
 This motivates us to impose the following boundary conditions which preserve the lower triangular form of the matrix $M$,
\begin{equation}\label{lower}
  \delta{M}_{\mu\nu} = \begin{pmatrix}
                 \mathcal{O}(e^{-4\phi}) &  \mathcal{O}(e^{-2\phi}) & \mathcal{O}(e^{-2\phi}) \\
                 \mathcal{O}(e^{-2\phi}) & \mathcal{O}(1) & \mathcal{O}(e^{-2\phi}) \\
                  \mathcal{O}(e^{-2\phi}) & \mathcal{O}(1) & \mathcal{O}(1)
               \end{pmatrix}.
\end{equation}
 with the order of coordinates $( \phi, \,u,\,v)$. Evidently, the Ba\~nabos metric \eqref{BTZ} satisfies the aforementioned boundary condition and is thus part of the phase space. In pure Einstein gravity in three dimensions, Brown-Henneaux boundary conditions also imply that the phase space is given by the Ba\~nanos metrics. It is interesting to see whether it is still true for the action \eqref{sugraaction}, which we leave for future work. To find the asymptotic symmetries, we need to find the allowed diffeomorphism $\xi^\mu$ and gauge transformation $\Lambda_\mu$ under which the variation 
\begin{equation}\label{dmuv}
	\begin{aligned}
		&\delta_{\xi,\Lambda} {M}_{\mu\nu} = \mathcal{L}_\xi {M}_{\mu\nu} + \p_\mu \Lambda_\nu - \p_\nu \Lambda_\mu  \\
		&=\begin{pmatrix}
                 2 \p_\phi \xi^\phi  &e^{2\phi} \p_\phi \xi^v +  \p_u \xi^\phi + \p_{[\phi} \Lambda_{u]}  & \p_v \xi^\phi + \p_{[\phi} \Lambda_{v]}  \\
                 \p_u \xi^\phi - \p_{[\phi} \Lambda_{u]}  & e^{2\phi} \p_u \xi^v & \p_{[u} \Lambda_{v]}\\ e^{2\phi} \p_\phi \xi^u +
                 \p_v \xi^\phi  - \p_{[\phi} \Lambda_{v]} & \quad e^{2\phi}( 2\xi^\phi+\p_v \xi^v + \p_u \xi^u)  - \p_{[u} \Lambda_{v]} & e^{2\phi} \p_v \xi^u
               \end{pmatrix}
	\end{aligned}
\end{equation}
falls off no slower than \eqref{lower}. 
 The falloff condition for the diagonal terms leads to
\begin{equation}\label{dia}
	\begin{aligned}
		\p_\phi \xi^\phi &= \mathcal{O}(e^{-4\phi}),\\
		\p_u \xi^v & = \mathcal{O}(e^{-2\phi}),\\
		\p_v \xi^u & = \mathcal{O}(e^{-2\phi}).
	\end{aligned}
\end{equation}
The $(u, v)$ component  of \eqref{dmuv} is $\p_u \Lambda_v - \p_v \Lambda_u \sim \mathcal{O}(e^{-2\phi})$, and then the  $(v, u)$ entry  of \eqref{dmuv} leads to $2\xi^\phi+\p_v \xi^v + \p_u \xi^u \sim \mathcal{O}(e^{-2\phi})$. 
The rest of the off-diagonal components further fix the form of the asymptotic Killing vector as 
\begin{equation}\label{ASK}
	\begin{aligned}
		\xi^u & = f (u) -{1\over 2}e^{-2\phi}  \bar f'' ({v})  + \mathcal{O}(e^{-4\phi}) ,\\
		\xi^v & =  \bar f ({v}) -{1\over 2} e^{-2\phi} f'' (u) + \mathcal{O}(e^{-4\phi}) ,\\
		\xi^{\phi} & = -\frac{1}{2}\z(  f' (u) +  \bar f' ({v}) \y)+ \mathcal{O}(e^{-4\phi}),
	\end{aligned}
\end{equation}
which is accompanied  by  a gauge transformation \begin{equation}\label{AG}
	\begin{aligned}
		\Lambda_\phi &= \frac{\ell^2}{2} \z(f' (u)-    \bar f' ({v})\y),\\
		\Lambda_u & \sim   \mathcal{O}(e^{-2\phi}),\\
		\Lambda_v & \sim  \mathcal{O}(e^{-2\phi}).
	\end{aligned}
\end{equation}
 Note that the non-vanishing component in $\Lambda_\phi$ is necessary to keep ${\delta M_{u \phi}},\,\delta M_{\phi v}\sim \mathcal{O}(e^{-2\phi})$, a consequence of the Fefferman-Graham gauge 
 in \eqref{lower}. In principle, the one-form $\Lambda$ can also contain an exact term. Such a term does not contribute to the $B-$field or affect the conserved charges and henceforth will be set to zero.

The asymptotic Killing vector has to preserve the boundary circle specified by \eqref{uvid}, which requires that the arbitrary functions $f(u)$ and $\bar{f}(v)$ parameterizing the asymptotic Killing vectors have to the periodic functions.  Under the Fourier expansion with
\begin{equation} f_n=e^{in u}, \quad  \bar{f}_n=-e^{-inv} , \label{modes}\end{equation} 
the asymptotic Killing vectors then form  a left and right-moving Witt algebra under the Lie bracket,
\begin{equation} [\xi_m,\,\xi_n]=-i(m-n)\xi_{m+n},\quad   [\bar \xi_m,\,\bar \xi_n]=-i(m-n)\bar \xi_{m+n} . \label{lie}\end{equation}
The conserved charges generating the asymptotic symmetries can be calculated in the covariant formalism \cite{Iyer:1994ys,Barnich:2007bf} from IIB supergravity \eqref{sugraaction}. As discussed in \cite{Compere:2007vx}, in general the covariant charges receive contributions from the Kalb-Romand field and the dilaton field in addition to the gravitational field. By explicit calculation using the expression of the charges in section 2 of \cite{Compere:2007vx}, however, we find that only the Einstein-Hilbert part of the action contributes to the charges of \eqref{ASK}\eqref{AG}, and therefore the calculation reduces to that in pure three-dimensional gravity \cite{Brown:1986nw}. 
As a result, the charges generating the transformation \eqref{ASK} and \eqref{AG} on the phase space with the boundary conditions \eqref{lower} are all finite, and  
they form left and right moving Virasoro algebras with central charges \begin{equation} 
c_L=c_R=6kN={3\ell\over 2G_{3}}.
\end{equation}

As a side remark, we comment on an alternative choice of boundary conditions. If the boundary conditions are relaxed to
\begin{equation}
  \delta{M}_{\mu\nu} = \begin{pmatrix}
                 \mathcal{O}(e^{-4\phi}) &  \mathcal{O}(1) & \mathcal{O}(1) \\
                 \mathcal{O}(1) & \mathcal{O}(1) & \mathcal{O}(e^{-2\phi}) \\
                  \mathcal{O}(1) & \mathcal{O}(1) & \mathcal{O}(1)
               \end{pmatrix},
\end{equation}
the gauge transformation \eqref{AG} is not necessary, and meanwhile terms of order $\mathcal{O}(e^{-2\phi})$ in the asymptotic Killing vector 
\eqref{ASK} can not be determined. Nevertheless, the covariant charges and the asymptotic symmetry algebra remain the same.  This is reminiscent of the story in pure Einstein gravity \cite{Porfyriadis:2010vg}.

To summarize, in this section, we study the asymptotic symmetry of IIB supergravity with the boundary conditions \eqref{lower}. Unlike pure Einstein gravity, the asymptotic Killing vectors \eqref{ASK} have to be accompanied by a gauge transformation \eqref{AG}. This is due to the presence of the non-vanishing Kalb-Ramond field $B_{\mu\nu}$, which plays a similar role as the $U(1)$ gauge field in the Einstein-Maxwell theory \cite{Hartman:2008dq} and the non-trivial $U(1)$ in AdS$_3$ with CSS boundary conditions  \cite{Compere:2013bya}.

\section{Asymptotic symmetry from the worldsheet: the Lagrangian formalism}
Asymptotic symmetry generators on the worldsheet were written in \cite{Giveon:1998ns} and further analyzed in \cite{deBoer:1998gyt,Kutasov:1999xu,Giveon:2001up}.  In this section, we revisit the problem with an emphasis on how to understand the spacetime asymptotic symmetries from the perspective of wordsheet theory. 
In the following, we will show that the asymptotic Killing vectors in the target spacetime can be found by requiring the worldsheet equation of motion to be preserved asymptotically. The worldsheet stress tensor is also found to be invariant under these transformations. This allows us to find the asymptotic symmetries of a general target spacetime from the worldsheet. 

\subsection{Spacetime boundary conditions from the worldsheet}
In this subsection, we elucidate how boundary conditions and the asymptotic Killing vectors on the target spacetime show up from the string worldsheet theory. The main idea is that the boundary condition for the target spacetime should be chosen such that the worldsheet equation of motion should be preserved in the large radius region. We will show that this requirement can be satisfied by the boundary conditions \eqref{lower}, from which we can further derive the asymptotic Killing vector \eqref{ASK} and gauge transformation \eqref{AG}.

The classical worldsheet action on a general string background in the conformal gauge can be written as 
\begin{equation}\label{Action:M}
  S=\frac{1}{2\pi \aaa'}\int dz^2 {M}_{\mu\nu} \p  { X}^\mu \bar \p { X}^\nu, \end{equation}
  where $M=g+B$ is the same combination defined in \eqref{Mdef}.
Then the equations of motion can be written as 
\begin{equation}
{\bar \p}(M_{\mu\nu} \p  X^\nu)+ \p(M_{\nu\mu} {\bar \p}  X^\nu)-\p_{ \mu} {M}_{\lambda\nu}\p  X^\lambda\bar{\partial}  X^\nu=0.\label{eom}
\end{equation}
From the worldsheet perspective, we would like to impose boundary conditions so that the worldsheet equation of motion \eqref{eom} all take the same form asymptotically, i.e. the difference from that of the zero-mass BTZ background vanishes, 
\begin{equation}
{\bar \p}(\delta M_{\mu\nu} \p  X^\nu)+ \p(\delta M_{\nu\mu} {\bar \p}  X^\nu)-\p_{ \mu} {\delta M}_{\lambda\nu}\p  X^\lambda\bar{\partial}  X^\nu=\mathcal{O}(e^{-2\phi}),\label{deom}
\end{equation}
where $\delta M$ is the deviation of $M$ from that of the massless BTZ \eqref{1}.
We note that the radial component in \eqref{deom} can easily be satisfied if we impose the Fefferman-Graham gauge \eqref{FG}, or equivalently if we use the first row and first column of the condition \eqref{lower}. 
For the $u$ and $v$ component of \eqref{deom}, one can check that all the terms vanishes after using $\bar \p u\to0, \, \p v\to0$, except for the two terms $\bar \partial (\delta M_{uu}\p u)$ and $ \partial (\delta M_{vv}\bar\p v)$. 
Finally, the action of \eqref{ASK} and \eqref{AG} on all the Ba\~nados metric \footnote{As mentioned in the last section, there might be solutions satisfying the boundary conditions \eqref{lower} beside the Ba\~nanos metrics \eqref{BTZ}. If so, it is interesting to check if the condition \eqref{additional} can be satisfied on these new solutions, which we leave for future study. } \eqref{BTZ} leads to \begin{equation}\p_v(\delta {M}_{uu})=\p_v(\delta { g}_{uu})\sim 0 ,\quad \p_u(\delta M_{vv})=\p_u(\delta {g}_{vv})\sim 0,\end{equation}
which after using the fact that $u,\,v$ are asymptotically chiral and anti-chiral indeed satisfies
\begin{equation} 
\bar \p \big(\delta M_{uu}\p u\big)=\bar \p \big(\delta  g_{uu}\p u\big)=\mathcal{o}(1), \quad  \p \big(\delta M_{vv} \bar \p v\big)=\p \big(\delta g_{vv} \bar \p v\big)=\mathcal{o}(1).\label{additional}
\end{equation}
Thus we have shown that the worldsheet equation of motion is preserved asymptotically \eqref{deom} by the boundary condition \eqref{lower}, which furthermore leads to the asymptotic Killing vector \eqref{ASK} and gauge transformation \eqref{AG}. 
In addition, it is not difficult to verify that the asymptotic Killing vectors  \eqref{ASK} supplemented by the gauge transformation \eqref{AG}
also leave the worldsheet action invariant,
\begin{equation}   \delta_{{\xi},{\Lambda}}{S}={1\over2\pi \aaa'}\int dz^2 \Big(\delta M_{\mu\nu} \p X^\mu\bar \p X^\nu
    \Big)  \overset{ \phi\rightarrow \infty}{\approx}0.\label{deltaS}
\end{equation}

To conclude, we find an interpretation of the boundary conditions \eqref{lower} discussed previously in the supergravity analysis: on string backgrounds that are asymptotically AdS$_3$, the boundary conditions \eqref{lower} imposed on spacetime fields
corresponds to the requirement that the worldsheet equations of motion are satisfied asymptotically \eqref{deom} at the AdS$_3$ boundary. 
The resulting asymptotic Killing vectors \eqref{ASK} and \eqref{AG} leave the worldsheet action invariant asymptotically, and hence indeed generate symmetry transformations on the worldsheet theory in the asymptotic region. 

\subsection{Asymptotic symmetry from a fixed background}\label{section:AS-AdS}
In the previous section, we 
considered a family of string worldsheet theories whose target spacetimes are asymptotically AdS$_3$, and showed that 
 the equations of motion of the worldsheet theory all take the same form in the region $\phi\to \infty$ \eqref{deom}.  
In this subsection, we focus on a given string background instead of a class of them. From the perspective of a two-dimensional quantum field theory, asymptotic symmetry can be obtained by looking for variations of the worldsheet fields which leaves the equations of motion invariant as $\phi\to \infty$. 
We will show that this requirement also leads to the asymptotic Killing vector \eqref{ASK}.  While the connection with the supergravity analysis is transparent in the previous subsection, considering a specific target spacetime enables us to construct Noether charges that generate the asymptotic symmetries.

The worldsheet equations of motion on the massless BTZ background \eqref{Action:AdS} can be written as
\begin{equation}\label{eom0}
    \begin{aligned}
        &\partial \bar{\partial} \phi-\exp(2 \phi)\bar{\partial}{u}\partial{v}=0,\\
        &\bar \partial {j}_0=k \bar{\partial}({\exp(2 \phi)\partial{v}})=0,\\
        &{ \partial} {\bar j}_0=k\partial({\exp(2 \phi)\bar{\partial}{u}})=0.
    \end{aligned}
\end{equation}
In the large radius region, the equations of motion imply that $u$ is approximatively chiral, ${u}$ is anti-chiral, and $ \phi$ is a free field. 
More explicitly,
we have \begin{equation}\label{chiralasy}\begin{aligned}
 &\bar \p u=\mathcal{O}(e^{-2\phi}),\quad \p v =\mathcal{O}(e^{-2\phi}) .
\end{aligned}
\end{equation}
This motivates us to propose the following condition: Under a diffeomorphism $\delta x^\mu=\xi^\mu$,  
 the chirality condition \eqref{chiralasy} and the equation of motion \eqref{eom0} are both preserved asymptotically, so that
\begin{equation}\label{eomasy-AdS}
    \begin{aligned}
        &\partial\bar{\partial}\xi^\phi-2\xi^\phi\exp(2\phi)\bar{\partial}u\partial v-\exp(2\phi)\bar{\partial}\xi^u\partial v-\exp(2\phi)\bar{\partial} u\partial\xi^v=\mathcal{O}(e^{-4\phi}),\\
        &\bar{\partial}\z(\exp(2\phi)\partial \xi^v+2\xi^\phi\exp(2\phi)\partial v\y)=\mathcal{O}(e^{-2\phi}),\\
        &\partial\z(\exp(2\phi)\bar{\partial}\xi^u+2\xi^\phi\exp(2\phi)\bar{\partial}u\y)=\mathcal{O}(e^{-2\phi}),\\
    \end{aligned}
\end{equation}
and
\begin{equation}\label{chiralasy2}
    \bar{\partial} {\xi}^u=\mathcal{O}(e^{-2\phi});\quad \partial{ \xi}^{v}=\mathcal{O}(e^{-2\phi}).
\end{equation}
The chirality condition \eqref{chiralasy2} can be solved asymptotically by  \begin{equation}\label{xiu0}
\begin{aligned}
            &    \xi^u=f(u)+\exp(-2\phi)A(u,v)+\mathcal{O}(e^{-4\phi}),\\
        &\xi^v=\bar{f}(v)+\exp(-2\phi)B(u,v)+\mathcal{O}(e^{-4\phi}),
\end{aligned}
\end{equation}
where $f,\,\bar f, \,A, \,B$ are arbitrary functions of their arguments. 
Then the second and the third equation of \eqref{eomasy-AdS} can be solved by \begin{equation}\label{xiphiAdS}
\begin{aligned}
        &{\xi}^\phi=-\frac{1}{2}f^\prime(u)-\frac{1}{2}{\bar f}^\prime({v})+\exp(-2\phi)D(u,v)+\mathcal{O}(e^{-4\phi}),\\
        &A=A(v),\quad B=B(u).
\end{aligned}
\end{equation}
Finally using the first equation of \eqref{xiu0}, we can solve for $A,\,B,\, D$ and get the asymptotic Killing vector \eqref{ASK}, 
which was previously derived by requiring that the boundary conditions \eqref{lower} from supergravity analysis.

On the other hand, we note that it is not clear how to find the gauge transformation \eqref{AG} from the equation of motion on a fixed background. On the other hand, \eqref{AG} is necessary if we further require that the action is invariant up to the leading order.
To see this, let us first write down the variation of the action under the left moving conformal transformation \eqref{ASK} alone, 
\begin{equation}
\begin{aligned} &\delta_{\xi_f}S=-\frac{k}{4\pi}\int d^2z\epsilon\z\{f^{\prime\prime\prime}(u)\bar{\partial} u\partial u+f''(u)(\partial u \bar\partial\phi-\partial \phi\bar\partial u)\y\}.
\end{aligned}
\end{equation}
The first and third terms both vanish after using the on-shell condition that $u$ is asymptotically chiral. The second term is of order $\mathcal{O}(1)$, but can be compensated by the additional gauge transformation \eqref{AG}. A similar discussion applies to the right-moving transformation parameterized by $\bar f(v)$ as well. 
Thus we have rederived the asymptotic Killing vector and gauge transformation from the string worldsheet theory on the massless BTZ background.

\subsection{The Noether charges}\label{sectNoether}
In the following, we will work out the asymptotic Noether charges that generate the transformation \eqref{ASK} and \eqref{AG}. In this subsection, we focus on the left-moving sector depending on $f(u)$ explicitly. A parallel discussion can be carried out for the right-moving sector as well.

Similar to the discussion for the classical action \eqref{Action:AdS}, the asymptotic Killing vector and gauge transformation for the quantum action \eqref{quantumaction} can be written as 
\begin{equation}
\begin{aligned}
        &\xi_f=f(u)\partial_u-\frac{k-2}{2k}\exp(-2\phi)f^{\prime\prime}(u)\partial_v-\frac{1}{2}f^\prime(u)\partial_\phi,\\
        &\Lambda_f=\frac{\ell^2-2\alpha'}{2} f'(u) d\phi.
\end{aligned}
\end{equation}
To find the Noether current, we introduce an arbitrary function  $\epsilon(z,\bar z)$ and consider the variation of the action \eqref{quantumaction} under the diffeomorphism and gauge transformation generated by $\epsilon \xi_f,\,\epsilon \Lambda_f$, 
\begin{equation}
\begin{aligned}    \delta_{\epsilon\xi_f,\epsilon\Lambda_f}S=\frac{1}{2\pi}\int d^2z\z\{\epsilon V_f+\partial \epsilon  j_{\bar{z}} +\bar{\partial}\epsilon j_z\y\},
\end{aligned}
\end{equation}
where 
\begin{equation}
\begin{aligned}\label{jV}
    &j_{z}=kf(u)\exp(2\phi)\partial v-(k-2)f^\prime(u)\partial\phi,\\
    &j_{\bar z}=-\frac{k-2}{2}f^{\prime\prime}(u)\bar{\partial}u,\\
    &V_f=-\frac{k-2}{2}f^{\prime\prime\prime}(u)\bar{\partial}u\partial u.
\end{aligned}
\end{equation}
We have the on-shell relation
\begin{equation}
    \bar{\partial}j_{z}+\partial j_{\bar z}=V_f.
\end{equation}
In the large radius region with $\phi\to\infty,$ both the antiholomorphic component of the current $j_z$ and the vertex $V_f$
are of order $\mathcal {O}(e^{-2\phi})$, and thus we are left with a holomorphic current $j_f\equiv j_z$ as expected. 
One can expand ${f}({u})$ in Fourier modes ${f}_n=\exp(in{u})$ as in \eqref{modes}, so that the Noether charges are given by
\begin{equation}\label{VirLAdS}
    \begin{aligned}
         {J}_{n}\equiv -\frac{1}{2\pi }\oint dz{j}_{n},  \quad  {j}_{n}\equiv   \exp(inu) j_0 - in(k-2)\exp(inu)\partial\phi.
    \end{aligned}
\end{equation}
Similarly, using the mode expansion ${\bar{f}}({v}) = -e^{-in {v}}$, the anti-holomorphic charges are given by 
\begin{equation}\label{VirRAdS}
    \begin{aligned}
        {\bar{J}}_{n}&\equiv -\frac{1}{2\pi } \oint d\bar{z}{\bar{j}}_{n},\quad 
        {\bar{j}}_{n}\equiv\exp(-in{v}){\bar{j}}_0+in(k-2)\exp(-in{v})\bar{\partial} \phi,
    \end{aligned}
\end{equation}
where $j_0$ is given by \eqref{tildej}.
The charges \eqref{VirLAdS} and \eqref{VirRAdS} are compatible with those obtained in \cite{Giveon:1998ns,deBoer:1998gyt}\footnote{More precisely, the non-chiral currents in \cite{deBoer:1998gyt} correspond to the Brown-Henneaux generators \eqref{ASK} without the gauge transformation \cite{Brown:1986nw}. The currents in \cite{Giveon:1998ns} are chiral and are consistent with \eqref{VirLAdS} up to conventions, although the appearance of the gauge transformation \eqref{AG} was not discussed either. }. The charges form left and right moving Virasoro algebras 
\begin{equation}
    \begin{aligned}
        \left[{J}_n,{J}_m\right]&=(n-m){J}_{n+m}+\frac{c}{12}n^3\delta_{n,-m},\\
\left[{\bar{J}}_n,{\bar{J}}_m\right]&=(n-m){\bar{J}}_{n+m}+\frac{\bar{c}}{12}n^3\delta_{n,-m},\\
        \left[{J}_n,{\bar{J}}_m\right]&=0,\\
    \end{aligned}\label{Viralg}
\end{equation}
where the central charges depend on the worldsheet topology and is given by 
\begin{equation}\label{central}
    c=\bar{c}=6k\mathcal{I},\quad \mathcal{I}= \frac{1}{2\pi} \oint dz\partial {u}=  w.
\end{equation}

Note that physical states in string theory also have to satisfy the Virasoro constraints. In order to check whether the spacetime Virasoro generators  \eqref{VirLAdS} and \eqref{VirRAdS} indeed generate asymptotic symmetries of string theory, we need to verify that they preserve the stress tensor. Using the free field OPE \eqref{opeAdS} in the asymptotic region, we find that all the worldsheet currents $j_m$ are primary operators with conformal weights $(1,0)$ as their OPE with the worldsheet stress tensor are given by  \begin{equation}\label{jT}
    \begin{aligned}
        & { T}_{ws}(z)  {j}_m(w)
       =\frac{{j}_m(w)}{(z-w)^2}+{\p j_m(w)\over z-w}+\cdots
    \end{aligned}
\end{equation}
Therefore $j_m$ are physical operators. 
{Using \eqref{jT},  we can also show that the spacetime Virasoro generators leave the stress tensor invariant,
\begin{equation}
[J_m, T_{ws}(w)]=[J_m, {\bar T}_{ws} (w)]=0 , \label{JmLn}
\end{equation}
which means that the constraints on the physical states are preserved by the spacetime Virasoro transformations. 

We have shown that the spacetime Virasoro generators leave the equation of motion asymptotically invariant, and furthermore preserve the Virasoro constraints. Therefore they generate symmetry transformations in the physical Hilbert space of string theory on asymptotically AdS$_3$ spacetime. The above analysis allows us to draw a general lesson about asymptotic symmetry from the string worldsheet, which can potentially be used for other string backgrounds. In section 6, we will apply this method to asymptotically flat spacetime and reproduce the BMS$_3$ symmetry at null infinity. In \cite{Du:2024xxx}, we will use the same strategy to study asymptotic symmetry for an asymptotically linear dilaton background that can be obtained by TsT transformations \cite{Apolo:2019zai}. 

 \subsubsection*{The Ward identity}
To see that the charges \eqref{VirLAdS} indeed implement the symmetry transformations, we need to study the Ward identity. In the derivation of the charges \eqref{VirLAdS}, we have used the fact that both the antiholomorphic current $j_{\bar z}$ and the vertex $V_f$ vanish at the boundary. As we will show momentarily, however, the vertex term plays an important role in the Ward identity even though it does not contribute to the expression of the charges.
By considering the variation of one point function $\delta \langle \mathcal{O}(z,\bar z)\rangle=\langle \delta \mathcal{O}(z,\bar z)\rangle-\langle \delta S \, \mathcal{O}(z,\bar z)\rangle=0$, we obtain the following Ward identity
\begin{equation}\label{ward}\begin{aligned}
    \delta_f\mathcal{O}(z,\bar{z})&=\frac{i}{2\pi}\oint_{\partial \Sigma} dw j_{w}\mathcal{O}(z,\bar{z})-\frac{i}{2\pi}\oint_{\partial \Sigma} d\bar{w}  j_{\bar{w}} \mathcal{O}(z,\bar{z})+\frac{1}{2\pi}\int_{\Sigma} d^2w V_f(w,\bar{w})\mathcal{O}(z,\bar{z}) ,
    \end{aligned}
\end{equation}
where $\Sigma$ is an open set with $(z,\bar{z})\in \Sigma$, and the expression of the currents and vertex are given by \eqref{jV}. The first term in \eqref{ward} is the commutator between the generator $J_f$ and the operator.
Using the OPE \eqref{opeAdS}, we learn that the nonvanishing part of the OPE $j_{\bar w}\mathcal{O}(z,\bar z)$ at large $\phi$ is proportional to $\delta^2(z-w)$. Therefore the second term in the above expression vanishes in this limit, as $(z,\bar z)\notin \p\Sigma$.  
On the other hand, the delta function contributes nontrivially in the last term of \eqref{ward} where the range of integration is over  $\Sigma$. 
Putting pieces together, we learn from \eqref{ward} that the variation of an arbitrary operator $\mathcal{O}$ under the symmetry transformation parameterized by $f$ is no longer given by the commutator but contains an extra term,
\begin{equation}\label{dO}
    \delta_f \mathcal{O} (z,\bar z)
    =-i[J_f, \mathcal{O}(z,\bar z)]-\frac{k-2}{2}f^{\prime\prime\prime}(u)\partial u \frac{\delta\mathcal{O}}{\delta j_0} (z,\bar{z}).
\end{equation}
The extra term above is generically non-zero with the exception that the operator only depends on the fundamental operators $x^\mu=u,\,v,\,\phi$, but not on their derivatives. In this case, the extra term vanishes and the charge $J_f$ indeed generates the corresponding symmetry transformation via the commutator. In particular, we have
\begin{equation}
    \delta_f x^\mu=-i[J_f, x^\mu], \quad \delta_{\bar f} x^\mu=-i[\bar J_{\bar f}, x^\mu].
\end{equation}
On the other hand, the second term in \eqref{dO} can be non-vanishing if the operator $\mathcal O$ depends on the current $j_0$, and hence the variation differs from the commutator. 
For instance, the commutators between the Virasoro generators and the zero mode currents $j_0$ are given by  \begin{equation}\label{Jfj0}
[J_f,j_0(z)] =-if^\prime(u)j_0+i(k-2) f^{\prime\prime}(u)\partial\phi,
\end{equation}
whereas the direct variation of the current is 
\begin{equation}
    \delta_f j_0=-i[J_f,j_0(z)] -\frac{k-2}{2}f^{\prime\prime\prime}(u)\partial u.\label{deltaj0}
\end{equation}
Similarly, we have \begin{equation}\begin{aligned}
&\delta_{\bar f} {\bar j}_0=-i[{\bar J}_{\bar f},{\bar j}_0(\bar z)] -\frac{k-2}{2}{\bar f}^{\prime\prime\prime}(v){\bar\partial} v, \\
&\delta_f \bar j_0=-i[J_f,\,\bar j_0]=\delta_{\bar f}  j_0=-i[{\bar J}_{\bar f},\,j_0]=0 .\end{aligned} \end{equation}
As is shown in the next section, the extra term is closely related to the integrability of the charges in the Hamiltonian formalism.

To recapitulate, in this section, we have studied spacetime asymptotic symmetries on AdS$_3$ from the perspective of worldsheet string theory. The boundary condition \eqref{lower} in the supergravity analysis is interpreted as the requirement that the worldsheet equations of motion should be satisfied asymptotically in the region where $\phi\to\infty$. More explicitly, given the falloff conditions \eqref{deom} and \eqref{chiralasy} on the worldsheet, we can derive the asymptotic Killing vector \eqref{ASK}. Using the Noether procedure, we can obtain the Noether charges \eqref{VirLAdS} and \eqref{VirRAdS}. As a consistency check, the charges leave the Virasoro constraints invariant as shown in \eqref{JmLn}, and hence preserve the physical Hilbert space. Finally, we also note that the Ward identity on a generic operator picks up an additional term as shown in \eqref{dO}.

\section{The Hamiltonian formalism}
In this section, we use the Hamiltonian formalism to derive the charges that generate the asymptotic symmetries.  

Let us first illustrate the main idea of the phase space method \cite{1987thyg.book..676C}. 
Consider a local field theory whose phase space is parameterized by $\{x^i,p_i\}\equiv \{q^I\}$, where $x^i$s are the generalized coordinates, and $p_i$s are the generalized momenta. Let $\omega = \frac{1}{2}\omega_{IJ} \delta q^I \wedge \delta q^J$ be the symplectic 2-form on $\mathrm{Span}\{q^I\}$. Then the Poisson bracket of two operators $P $ and $Q$ is defined as $\{P,Q\} = \omega^{IJ} \p_I P \p_J Q$, where $\omega^{IJ}$ is the inverse of $\omega_{IJ}$. 
Consider a variation along an arbitrary vector in the phase space of the form \begin{equation} i_{\xi}  = \xi^I \frac{\delta}{\delta q^I},\end{equation}  
 where the variation $\xi^I\equiv \delta q^I$ includes variations of all the coordinates in the phase space, i.e. the variation of the generalized coordinates as well as the generalized momenta. 
 If the action of $\xi^I$ on an operator $P$ is generated by a charge $H_{\xi}$ via Poisson bracket, 
\begin{equation}\label{deltaP}
\delta_{\xi} P\equiv  \xi^J  \frac{\delta P}{\delta q^J}=\{H_\xi,P\} = \omega^{IJ} \frac{\delta H_\xi}{\delta q^I}\frac{\delta P}{\delta q^J} ,
\end{equation}
we will have 
\begin{equation}
	\xi^J  = \omega^{IJ} \frac{\delta H_\xi}{\delta q^I}.
\end{equation}
Then we learn that the infinitesimal charge is given by
\begin{equation}
\delta H_\xi\equiv   \frac{\delta H}{\delta q^I} \delta \Phi^I = \xi^K \omega_{KJ} \delta q^J,\label{inficharge}
\end{equation}
which is defined around a point in the phase space. 
 Finite charges can then be obtained by integrating  \eqref{inficharge} provided that the latter is integrable. 
If the variation $\xi$ depends on the phase space variables $q^I$, it is important to check if the infinitesimal charge \eqref{inficharge} is integrable.

For exact symmetry of the theory, 
it is natural to obtain the variations of momenta $\delta p_i$ from those of coordinates $\delta q^i$ by using the on-shell condition which relates the coordinates and momenta.  
On the other hand, the main purpose of this paper is to investigate asymptotic symmetry which preserves the equations of motion only approximately. This means that it is possible to consider $\delta q$ and $\delta p$ separately, with the requirement that equations of motion are satisfied up to a certain order. It is then possible to choose $\delta p$ in such a way that the corresponding charges satisfy several desirable properties including integrability and conservation. In the following, we carry out the explicit calculation in the example of AdS$_3$ string theory.

\subsection{Asymptotic charges for string theory on AdS$_3$}
 To write the Poisson brackets, let us first write the action \eqref{Action:AdS} for AdS$_3$ strings on Lorenzian worldsheet,
\begin{equation}
    \begin{gathered}
        S=-\frac{k}{4\pi}\int dt d\sigma\z\{\z(\p_\sigma\phi\y)^2-\z(\p_t\phi\y)^2+\exp(2\phi)\z(\p_\sigma u-\p_t u\y)\z(\p_\sigma v+\p_t v\y)\y\},
    \end{gathered}
\end{equation}
from which we obtain the canonical momentum \footnote{We have absorbed a factor of $2\pi$ into the definition of the momentum.}
\begin{equation}\label{pj}
     p_{\phi}\equiv 2\pi  {\delta S\over \delta (\p_t\phi)}=k\partial_t\phi,\quad p_u=j_0=\frac{k}{2}\exp(2\phi)(\partial_\sigma v+\partial_t v),\quad  p_v=-\bar{j}_0=-\frac{k}{2}\exp(2\phi)(\partial_\sigma u-\partial_t u),
\end{equation}
and the symplectic form, 
\begin{equation}\label{symp}
    \omega={1\over2\pi}\oint d\sigma (\delta x^\mu \wedge \delta p_\mu) . 
\end{equation}
Or equivalently, the Poisson bracket on the equal time slice is 
\begin{equation}\{ x^\mu(\sigma),\,p_\mu(\sigma')\}=2\pi \delta(\sigma-\sigma').
\end{equation}
The Hamiltonian is given by 
\begin{equation}
    H=\frac{1}{2\pi}\int d\sigma\z\{\frac{p_\phi^2}{2k}+\frac{k}{2}({\partial_\sigma\phi})^2+p_u\partial_\sigma u-p_v\partial_\sigma v+\frac{2}{ k}e^{-2\phi}p_up_v\y\},
\end{equation}
and the worldsheet energy-momentum tensor is
\begin{equation}\label{Tps}
    \begin{aligned}
        &T_{ws}=-\frac{1}{4k}\z((k\partial_\sigma\phi+ p_\phi)^2+4(k\partial_\sigma u+ \exp(-2\phi)p_v)p_u\y),\\
        &\bar{T}_{ws}=-\frac{1}{4k}\z((k\partial_\sigma\phi- p_\phi)^2-4(k\partial_\sigma v-\exp(-2\phi)p_u)p_v\y).
    \end{aligned}
\end{equation}

For string theory on AdS$_3$, the coordinates on the phase space can be chosen as $\{q^I(\sigma)\}=\{ u,v, \phi;p_u,p_v,p_\phi \}$.
 Let $H_f$ denote the charge which generates the transformation \eqref{xix} in the phase space $\{\xi^I\}=
\{\xi^u,\,\xi^v,\,\xi^\phi;\xi^{p_u},\,\xi^{p_v},\,\xi^{p_\phi}\}$. 
As discussed in the previous section,  the asymptotic Killing vectors \eqref{ASK} for string theory on AdS$_3$ spacetime are parameterized by two functions $f(u)$ and $\bar f(v)$. In the following, we focus on the left moving part with $\bar f(u)=0$,  the generator of
 which we reproduce here for convenience,  
\begin{equation}\label{xix}
	\xi^\phi=-\frac{1}{2}f^\prime(u),\quad \xi^u=f(u),\quad \xi^v =-\frac{1}{2}\exp(-2\phi)f^{\prime\prime}(u),
\end{equation}
where we have omitted all subleading terms in the large $\phi$ expansion. 
The discussion of the right moving part is similar.
The transformation \eqref{xix} is accompanied by a transformation of the conjugate momentum in the phase space $\xi^{p_u},\,\xi^{p_v},\,\xi^{p_\phi}$, which we now specify. 
For asymptotic symmetries,  the transformation has to preserve the Virasoro constraint as well as
 the equations of motion in the large $\phi$ region of the target spacetime. The condition that the transformation preserves the Virasoro constraint requires that the Poisson bracket between the charge $H_f$ and the Hamiltonian vanishes
 \begin{equation}\label{deltaH}
\begin{aligned}
	\delta_\xi H\equiv\{H_\xi,H\}= &\frac{1}{2\pi}\oint d\sigma \Bigg[   \Big(f'(u) p_u  - \frac{k}{2} f''(u) \p_\sigma \phi\Big)\Big(\p_\sigma u+ \frac{2}{ k} e^{-2\phi} p_v \Big) + \frac{1}{2} e^{-2\phi} f'''(u)  p_v \p_\sigma u \\
	 &+  \xi^{p_u}  \z(  \p_\sigma u + \frac{2}{ k} e^{-2\phi} p_v \y)  -  \xi^{p_v}  \z(  \p_\sigma v - \frac{2}{k} e^{-2\phi} p_u \y) + \xi^{p_{\phi}} \frac{p_\phi}{ k}  \Bigg] =\mathcal{O}(e^{-2\phi}) .
\end{aligned}
\end{equation}
To preserve the equations of motion, we require the Jacobi identity between $H,\, J_f$ and $\{q^I\}$ to be satisfied,
\begin{equation}\label{Jacobi}
0
=\{H_f,\,\{H,\,q^I \}\}+\{\xi^I,\,H \}+\{q^I,\,\{H_f,\, H\} \}.
 \end{equation}
Assuming that $(\xi^{p_u},\,\xi^{p_v},\,\xi^{p_\phi})$ are functionals of the phase space coordinates $q^I$, we can then solve \eqref{deltaH} and \eqref{Jacobi} perturbatively.

At the order of $\mathcal{O}(1)$, the Hamiltonian is preserved \eqref{deltaH} and hence the third term in \eqref{Jacobi} vanishes. 
Taking $q^I=\phi$ in \eqref{Jacobi}, 
 we get   
 \begin{equation}\label{xipphi}  \xi^{p_\phi}\equiv  \{H_f,\,p_\phi\}=-k\{\xi^\phi,\,H\}+\mathcal{O}(e^{-2\phi})
 = -\frac{k}{2} \p_\sigma u+\mathcal{O}(e^{-2\phi}).
\end{equation}
Similarly, applying \eqref{Jacobi} to $j_0,\bar j_0$, we get 
\begin{equation}\begin{aligned}
    \{H,\xi^{p_u}\}&=\{H_f,\{H,p_u \}\}+\mathcal{O}(e^{-2\phi})=-\partial_{\sigma} \xi^{p_u}+\mathcal{O}(e^{-2\phi}),\\
    \{H,\xi^{p_v}\}&=\{H_f,\{H,p_v\}\}+\mathcal{O}(e^{-2\phi})=\partial_\sigma \xi^{p_v}+\mathcal{O}(e^{-2\phi}),
\end{aligned}\end{equation}
which means that $\xi^{p_u}$ is holomorphic, and $\xi^{p_v}$ is antiholomorphic at the leading order.  
Then it is reasonable to assume that $\xi^{p_v}$ is independent of $u$, as the latter is holomorphic. Similarly, $\xi^{p_u}$ is independent of $v$. 
Plugging \eqref{xipphi} into \eqref{deltaH} and collecting terms proportional to $\p_\sigma u$ and $\p_\sigma v$, we get the leading order variation of the currents, 
\begin{equation}
    \begin{aligned}
        &\xi^{p_u} = - f ' (u) p_u + \frac{k}{2}f''(u) \p_\sigma \phi + \frac{1}{2}f''(u)p_\phi+\mathcal{O}(e^{-2\phi}),\\   &\xi^{p_v}=0+\mathcal{O}(e^{-2\phi}).
    \end{aligned}
\end{equation}
The above equation and \eqref{xipphi} constitute a solution of \eqref{deltaH} and \eqref{Jacobi} to the leading order. 

Now we proceed to the next leading order.  
Note that the Hamilton equation $\p_t v=\{v,H\}$ sets up a relation $\p v =\frac{1}{k}e^{-2\phi}p_u$, to preserve which we need to require the Jacobi \eqref{Jacobi} in the $v$ direction to be satisfied up to the order of $\mathcal{o}(e^{-2\phi})$. Then we get
\begin{equation}
    \frac{1}{2\pi}\oint d\sigma\z(\{v, \xi^{p_u}_{(2)}\}\,\p_\sigma u-\{v, \xi^{p_v}_{(2)} \}\,\p_\sigma v+\{v, \xi^{p_\phi}_{(2)} \}{p_\phi\over k}\y)-{1\over2}e^{-2\phi}f'''(u)\p_\sigma u+e^{-2\phi}f''(u){p_\phi\over k}  =0,\label{Jacobiv}
\end{equation}
where $\xi^{p_\mu}_{(2)}$ are terms in $\xi^{p_\mu}$ that are of order $\mathcal{O}(e^{-2\phi})$.
Similarly, taking $q^I=u$, we get 
\begin{equation}\label{Jacobiu}
    \frac{1}{2\pi}\oint d\sigma\z\{\{u, \xi^{p_u}_{(2)}\}\,\p_\sigma u-\{u, \xi^{p_v}_{(2)} \}\,\p_\sigma v+\{u, \xi^{p_\phi}_{(2)} \}{p_\phi\over k}\y\}  =0.
\end{equation}
The simplest solution to the above two equations is to take the ansatz $e^{2\phi}\xi^{p_\mu}=A^\mu(u,v,\phi) p_v$, so that the solution \footnote{Note that \eqref{xip} is not the unique solution to \eqref{Jacobiu} and \eqref{Jacobiv}. For instance, one can add terms like $e^{-2\phi}B^\mu (u,v,\phi) p_\phi+C^\mu(u,v,\phi)$ to the solution \eqref{xip} and construct a new solution. However, adding these terms will in general make the infinitesimal charge \eqref{inficharge} non-integrable. } is 
\begin{equation}\begin{aligned}\label{xip}
\xi^{p_u} &= - f'(u) p_u+ \frac{k}{2}f''(u) \p_\sigma \phi + \frac{1}{2}f''(u)p_\phi + \frac{1}{2} e^{-2\phi} f'''(u)  p_v,\\
\xi^{p_v}&= 0,\\
\xi^{p_\phi}&= -{k\over 2} (\p_\sigma u+\frac{2}{ k} e^{-2\phi} p_v)f''(u).
\end{aligned}
\end{equation} 
 Plugging the variation \eqref{xix} and \eqref{xip} into the expression \eqref{inficharge}, we obtain the infinitesimal charge 
 \begin{equation}\label{dH}
    \begin{gathered}
        \delta H_f=\frac{1}{2\pi}\oint d\sigma\delta\z(-f(u)p_u+\frac{1}{2}f^{\prime}(u)(k\partial_\sigma \phi+ p_\phi)+\frac{k}{2}f^{\prime\prime}(u)e^{-2\phi}p_v\y),
    \end{gathered}
\end{equation}
which is integrable and the resulting finite charge is given by
 \begin{equation}\label{hf}
 H_f =\frac{1}{2\pi}\oint d\sigma \z(-f(u)p_u+\frac{1}{2}f^{\prime}(u)(k\partial_\sigma \phi+p_\phi)+\frac{k}{2}f^{\prime\prime}(u)e^{-2\phi}p_v\y).
 \end{equation}
As a consistency check, we can verify that the charge \eqref{hf} indeed generates the transformation \eqref{xix} and \eqref{xip} in the phase space via the Poisson bracket, namely \begin{equation}
    \{H_f,\, q^I\}=\xi^I.\label{Hvari}
\end{equation}  
Note that the last term in \eqref{hf} is of order $\mathcal{O}(e^{-2\phi})$. We have kept the term so that it can generate the transformation $\xi^v$ \eqref{xix} which is of the same order. This subleading term does not play a role in many calculations and can be ignored with a few exceptions. In particular, the classical charge \eqref{hf} ignoring the last term agrees with the Noether charge derived in the Lagrangian formalism \eqref{VirLAdS} in the large $k$ limit.
Then the Poisson bracket \eqref{Hvari} 
 agrees with \eqref{Jfj0} in the Lagrangian formalism after the replacement $i\{,\,\}\to[,\,]$.
The worldsheet energy-momentum tensor \eqref{Tps}  are also invariant under the transformation \eqref{xix} and \eqref{xip}, 
\begin{equation}\label{HTws}
    \{H_f,T_{ws}(\sigma)\}\overset{\phi\rightarrow\infty}{\approx}0,\quad
    \{H_f,\bar{T}_{ws}(\sigma)\}\overset{\phi\rightarrow\infty}{\approx}0 ,
\end{equation}
which means that the Virasora Constraints are preserved at the classical level.
Furthermore, the Fourier modes $H_{f_n}\equiv H_n$ form an Virasoro algebra
\begin{equation}\label{HLalg}
    i\{H_{n},H_{m}\}=(n-m)H_{n+m}+\frac{n^3c}{12}\delta_{n,-m}.
\end{equation}
The conservation of stress tensor \eqref{HTws} and algebra \eqref{HLalg} are just the classical version of \eqref{JmLn} and \eqref{Viralg}.

 To end this section, let us comment again on the subtly regarding the variation of canonical momentum that first appeared in \eqref{deltaj0}. We have seen in the Lagrangian formalism that direct variation of the current $\delta_f j_0$ differs from acting the charge $J_f$ on it via commutator. 
In the Hamiltonian formalism, however, variations agree with the Poisson bracket by definition \eqref{deltaP}. 
As can be checked explicitly, the variations \eqref{xip} derived in the Hamiltonian formalism actually agree with the commutator \eqref{Jfj0}, and differ from the direct variation in Lagrangian formalism by a term proportional to $f'''(u)$ \eqref{deltaj0}.
The reason for the difference is that
in the phase space variations of the coordinates and their conjugate momentum can be performed independently, without assuming their relation \eqref{pj} a prior. 
We then determine the variation of canonical momentum by requiring the Virasoro constraint as well as the equation of motion to be satisfied asymptotically. The resulting $\xi^{p_\mu}$ can thus be different from direct variation \eqref{deltaj0} which implicitly assumes that the relation \eqref{pj} is satisfied to all orders.
As we have shown explicitly, the variation \eqref{xip} leads to an integrable charge \eqref{hf}. Instead, if we use \eqref{deltaj0}, i.e. the expression obtained by directly varying \eqref{pj}, the infinitesimal charge will contain the integrable part given by \eqref{dH} and an additional non-integrable part. Nevertheless, the integrable part of the later approach is also given by \eqref{hf}, which also agrees with the Noether charge \eqref{VirLAdS} at large $\phi$ and large $k$.  
Therefore the analyses in the Lagrangian formalism and Hamiltonian formalism are compatible with each other.

\section{Flat spacetime in three dimensions}
In this section, we apply the method to the worldsheet theory of flat spacetime in three dimensions, the latter of which can be regarded as the bosonic sector of superstring theory in ten dimensions with a seven-dimensional compact internal space. By requiring the worldsheet equations of motion and constraints to be preserved asymptotically near $\mathcal I^+$, we find the BMS$_3$ transformations and write down the Noether charges in both the Lagrangian and Hamiltonian formalism. In addition, we find that a dilatation transformation also preserves the equations of motion asymptotically, although it does not preserve the worldsheet action or Hamiltonian.  

\subsection{String worldsheet theory in three dimensions}
Motivated by the appearance of BMS$_3$ symmetry in Einstein gravity \cite{Barnich:2006av, Barnich:2012aw, Barnich:2010eb}, we write the metric of flat spacetime in
the Bondi gauge
\begin{equation}\label{bondi}
    ds^2=-du^2-2dudr+r^2d\theta^2,\quad \theta\sim \theta+2\pi.
\end{equation}
where null future $\mathcal{I}^+$ is at $r\to\infty$.
The coordinate transformation from \eqref{bondi} 
to Cartesian coordinates is given by 
\begin{equation}\label{cdtrans}
     x^0=u+r, \quad x^1=r \cos\theta, \quad x^2=r\sin\theta.
\end{equation}
The worldsheet theory on the target spacetime \eqref{bondi} is given by the action
\begin{equation}\label{actionflat}
    S=\frac{1}{2\pi \alpha^\prime}\int dz^2 \z\{-\partial u \bar{\partial} u-\partial u\bar{\partial}r-\partial r\bar{\partial}u+r^2\partial\theta\bar{\partial}\theta\y\},
\end{equation}
with the stress tensor  
\begin{equation}
    \begin{aligned}\label{Tflat}
        &T_{ws}=\frac{1}{2\pi\alpha^\prime}\z\{-\partial u\partial u -2\partial u\partial r +r^2\partial\theta\partial\theta\y\},\\
        &\bar{T}_{ws}=\frac{1}{2\pi\alpha^\prime}\z\{-\bp u\bp u -2\bp u\bp r +r^2\bp\theta\bp\theta\y\}.
    \end{aligned}
\end{equation}
The equations of motion are given by 
\begin{equation}\label{eomflat}
    \begin{aligned}
        &\partial\bar{\partial}(u+r)=0,\\
        &\partial\bar{\partial}\theta+\partial \log r\bar{\partial}\theta+\bar{\partial}\log r\partial\theta=0,\\
        &\partial\bar{\partial}u+r\partial\theta\bar{\partial}\theta=0.
    \end{aligned}
\end{equation}
General solutions to the equations of motion \eqref{eomflat} can be obtained from the mode expansion in Cartesian coordinates  
\begin{equation}
x^\mu=x^\mu_0+{p^\mu }t+\sum_{n\in \mathbb Z^+} x^\mu_n. \label{3dsol}\end{equation} 
where $x^\mu_n$ are oscillating modes \begin{equation}  x^\mu_n\equiv \z(\alpha^\mu_n   e^{in \sigma^+}+ \tilde \alpha^\mu_n e^{-in \sigma^-}\y)+c.c.
, \quad
\alpha^\mu_n=(\alpha^\mu)^\dagger_{-n},
\end{equation}
where $\sigma^\pm=\sigma\pm t$.
Physical states at the classical level are obtained by further requiring the full stress tensor including the internal dimensions vanishes. 
As we are interested in modes that propagate to the null future, it is sufficient to consider the massless modes, which correspond to excitations at level $1$ either in three non-compact dimensions or in the internal dimensions. 
 Now we focus on the case with a level $1$ excitation in both the left and right moving sector of the solution \eqref{3dsol}. The constraints at this level are then the massless condition and the polarization conditions, 
\begin{equation}
p^2=0,\quad p_\mu \alpha_{-1}^\mu=p_\mu \alpha_{1}^\mu=p_\mu \tilde\alpha_{-1}^\mu=p_\mu \tilde\alpha_{1}^\mu=0.\label{onshell}
\end{equation}
The second condition above implies that \begin{equation}p_\mu x_{1}^\mu=0.\label{onshell2}\end{equation}
Massless modes in the Bondi gauge can then be obtained by plugging the mode expansion \eqref{3dsol} into the coordinate transformation \eqref{cdtrans}. 
The resulting expression contains a square root and is very complicated in the general case.
For our purposes, we are interested in outgoing states, which reach $\mathcal I^+$ at late time. This enables us to expand the 
massless string solutions in Bondi gauge at large $t$
\begin{equation}
    \begin{aligned}
        &r=r_0+p^0 t+\frac{p_i x_{1}^i}{p^T}+\mathcal{O}(t^{-1}),\quad i=1,2\\
        &u=u_0-\frac{p_\mu x_{1}^\mu}{p^T}+\mathcal{O}(t^{-1})=u_0+\mathcal{O}(t^{-1}),\quad {\mu=0,1,2}\\
        &\theta=\arctan(\frac{p^2}{p^1})+\mathcal{O}(t^{-1}).
    \end{aligned}\label{largersol}
\end{equation}
where $r_0,\,u_0$ are constants, and we have used the constraint \eqref{onshell2} in the second expression for $u$. 
The above expansion is consistent with the assumption of large $t$ and large $r$, as they are of the same order. We thus conclude that the massless solutions have the following falloff behavior as $r\to \infty$, 
\begin{equation}
    \theta \sim u\sim \mathcal{O}(1).
        \label{falloff}
\end{equation}
Further taking derivatives of \eqref{largersol} we obtain the following falloff 
   \begin{equation}    \label{falloff2} 
    \begin{aligned} &\p r\sim \bp r\sim \mathcal{O}(1),\quad \p \theta\sim\bp \theta\sim\p u\sim\bp u\sim \mathcal{O}(r^{-1}),\\ 
        &\p\bp u\sim \p\bp r\sim \mathcal{O}(r^{-1}),\quad \p\bp \theta\sim \mathcal{O}(r^{-2}).
        \end{aligned}\end{equation}
Note that the density of the spacetime angular momentum is divergent as $r^2\p \theta \sim t { x_1^\mu}+\mathcal{O}(1)$. However, the divergent term is a periodic function and hence its integral over $\sigma$ vanishes. Therefore the spacetime angular momentum is finite
\begin{equation}
L={1\over 2\pi \alpha'}\int d\sigma~ r^2 \p_t \theta \sim \mathcal{O}(1).\label{falloffam}
\end{equation} 
So far we have been discussing massless modes with both the left and right excitations in the non-compact three dimensions. There are two other possibilities for massless excitations of closed strings. One is that both the left and right moving excitations are in the internal space, which is nothing but massless particles in three dimensions, and the solution is given by null geodesics. The third type of massless mode is that excitation in one sector is in flat space and the other is in the internal space. It is easy to verify that all these cases satisfy the falloff condition \eqref{falloff}, \eqref{falloff2}, and \eqref{falloffam}.

\subsection{BMS$_3$ generators from the worldsheet}

Now consider a spacetime diffeomorphism $\delta x^\mu=\xi^\mu$ that preserves the solution of string theory asymptotically. It is natural to assume that the variation $\xi^\mu$ also preserves the falloff condition \eqref{falloff}, so that $\xi^\mu$ can be expanded at large $r$ as
\begin{equation}\begin{aligned}  \label{xiexp} \xi^{u}&=\sum_{n=0} r^{-n}\xi^u_{(n)}(\theta,\,u),\\
\xi^\theta&=\sum_{n=0} r^{-n}\xi^\theta_{(n)}(\theta,\,u),\\
\xi^r&=\sum_{n=-1} r^{-n}\xi^r_{(n)}(\theta,\,u).
\end{aligned}\end{equation}
In addition, the equations of motion \eqref{eomflat} should also be satisfied asymptotically. In the large radius expansion, we impose the following falloff condition on the equations of motion
\begin{equation}\label{eomfalloff}
    \begin{aligned}
        &\partial\bar{\partial}(\xi^u+\xi^r)=\mathcal{O}(r^{-1}),\\        &\p\bp\xi^\theta+{1\over r }\big(\p\xi^{r}\bp\theta+\bp\xi^{r}\p\theta+\p r\bp\xi^\theta+\bp r\p \xi^\theta\big)-\frac{\xi^r}{r^2}(\p r\bp\theta+\bp r\p\theta)=\mathcal{O}(r^{-4}),\\
        &\partial \bar{\partial}\xi^u+\xi^r\partial\theta\bar{\partial}\theta+r\bar{\partial}\theta\partial \xi^{\theta}+r\partial\theta\bar{\partial}\xi^{\theta}=\mathcal{O}(r^{-3}).
    \end{aligned}
    \end{equation}
The first equation in \eqref{eomfalloff} is automatically satisfied by the expansion \eqref{xiexp}.  
From the second equation, we have
\begin{equation}\label{soleq2}
    \begin{aligned}
        &\mathcal{O}(r^{-1}): \quad  \p_u\xi^\theta_{(0)}=0,\\
        &\mathcal{O}(r^{-2}): \quad  \p_u \xi^\theta_{(1)}-\p_\theta^2 \xi^\theta_{(0)}-2\p_\theta \xi^r_{(-1)}=0,\\
        &\mathcal{O}(r^{-3}): \quad \p_u^2 \xi^\theta_{(1)}=0;\quad \p_\theta \xi^\theta_{(1)}+\xi^r_{(0)}=0,
    \end{aligned}
\end{equation}
where we have omitted the arguments for brevity.
From the third equation in \eqref{eomfalloff}, we have
\begin{equation}
\begin{aligned}
        &\mathcal{O}(r^{-1}):\quad \p_u \xi^u_{(0)}-2\p_\theta \xi^\theta_{(0)}-\xi^r_{(-1)}=0,\\
        &\mathcal{O}(r^{-2}):\quad   \p_\theta \xi^u_{(0)}+\xi^\theta_{(1)}=0,\quad  \p_u^2 \xi^u_{(0)}=0, \quad \xi^u_{(1)}=0,
    \end{aligned}
    \end{equation}
which combined with \eqref{soleq2}
leads to the solution of the asymptotic Killing vectors

\begin{equation}
    \begin{aligned}
&\xi^u_{(0)}=F(\theta)+u(Y'(\theta)+d_0),\quad \xi^u_{(1)}=0,\\
        &\xi^\theta_{(0)}=Y(\theta),\quad \xi^\theta_{(1)}= -\frac{uY''(\theta)+F'(\theta)}{r},\\
        &\xi^r_{(-1)}=-r(Y'(\theta)-d_0),\quad \xi^r_{(0)}=F''(\theta)+uY'''(\theta),
    \end{aligned}\label{bms}
\end{equation}
where $F(\theta)$ and $Y(\theta)$ are both arbitrary periodic functions of $\theta$, and
$d_0$ is constant. 
The $d_0=0$ part of the asymptotic Killing vectors is given by 
\begin{equation}
    \begin{aligned}
&\xi^u=F(\theta)+u Y'(\theta)+\mathcal{O}(r^{-2}),\\
        &\xi^\theta=Y(\theta)-\frac{uY''(\theta)+F'(\theta)}{r}+\mathcal{O}(r^{-2}),\\
        &\xi^r=-r Y'(\theta)+F''(\theta)+uY'''+\mathcal{O}(r^{-1}),
    \end{aligned}\label{bms3}
\end{equation}
which are just generators of BMS$_3$, the asymptotic symmetry group of Einstein gravity on asymptotically flat spacetime with Bondi gauge \cite{Barnich:2010eb}. 
$F(\theta)$ parameterizes supertranslations, and $Y(\theta)$ parameterizes superrotations.

The vector proportional to $d_0$ is a dilatation  \begin{equation} \label{D}
D\equiv u\p_u+ r\p_r, \end{equation}
which is not an isometry of the metric \eqref{bondi}, but a conformal Killing vector. 
The appearance of the dilatation generator is reminiscent of the conformal BMS group discussed by \cite{Haco:2017ekf}. One natural question is whether we can also find the so-called BMS dilatation $E$  and BMS special conformal transformation $C$ found in \cite{Haco:2017ekf}. It turns out that both $E$ and $C$ require $\xi^r$ to have a growth of order $r^2$ at $\mathcal I^+$ and are hence not included under our choice of boundary conditions on the worldsheet \eqref{xiexp}. One may wonder if we can relax the boundary conditions \eqref{xiexp} to allow more additional conformal Killing vectors. This doesn't seem to be possible without spoiling the equations of motion \eqref{eomfalloff} asymptotically. This is understandable as we do expect conformal Killing vectors to modify the spacetime metric and hence the worldsheet equation of motion at $\mathcal{I}^+$. 
Nevertheless, the dilatation $D$ does preserve the worldsheet equation of motion as shown explicitly.

\subsection{Noether charges in the Lagrangian formalism}
Now let us work out the Noether charge on the worldsheet which generates the diffeomorphism \eqref{bms3} in the Lagrangian formalism. Let us first consider the supertranslation transformation $\xi_F$ parameterized by $F(\theta)$. The variation of the action \eqref{actionflat} generated by $\epsilon(z,\bar z) \xi_F$ is given by  
\begin{equation}
    \begin{aligned}
        \delta_{\epsilon\xi_{F}}S=&-\frac{1}{2\pi \alpha'}\int d^2z\z\{\p\epsilon((F+F'')\bp u+F\bp r+rF'\bp\theta)+\bp\epsilon((F+F'')\p u+F\p r+rF'\p\theta)\y\}\\
        &-\frac{1}{2\pi \alpha^\prime}\int d^2z~\epsilon(F^\prime+F^{\prime\prime\prime})(\partial u\bar{\partial}\theta+\partial\theta\bar{\partial} u).
    \end{aligned}
\end{equation}
Using the on-shell falloff condition \eqref{falloff2}, the vertex in the second line above is of order  $\mathcal{O}(r^{-2})$, and thus the action is asymptotically invariant. This allows us to write down the Noether charge \begin{equation}
    \begin{aligned}
        &H_F=
        \frac{1}{2\pi \alpha'}\oint d\sigma \z\{ F(\partial_t(u+r))+F''\p_t u+rF'\p_t\theta \y\}.\\
    \end{aligned}\label{HFL}
\end{equation}
Similarly, we obtain the Noether charge for superroations as 
\begin{equation}
    \begin{aligned}\label{HYL}
        &H_Y
        =\frac{1}{2\pi\alpha'}\oint d\sigma \z\{ (-rY'+uY'+uY''')\p_t u+uY'\p_t r+(urY''-r^2Y)\p_t\theta \y\}.
    \end{aligned}
\end{equation} 
From the on-shell falloff \eqref{falloff} and \eqref{falloff2}, it is not difficult to see that the charges for supertranslation \eqref{HFL} are always finite. On the other hand, the last term in \eqref{HYL} is apparently divergent as the integrand $r^2 Y\p_t \theta$ is of order $\mathcal{O}(r).$ Nevertheless, we note that $\theta\sim \theta_0$ from the solution \eqref{3dsol}, so that the potentially divergent term in \eqref{HYL} is proportional to $Y(\theta_0) L$, where $L$ the angular momentum \eqref{falloffam}. As argued before, the angular momentum is finite, and hence there is no divergence in the charge \eqref{HYL}.

Finally, let us consider the variation of the action \eqref{actionflat} generated by $\epsilon(z,\bar z) {D}$
\begin{equation}\label{dDS}
    \begin{aligned}
        \delta_{\epsilon D}S = & -\frac{1}{2\pi \aaa'} \int d^2 z \Big\{  \p \epsilon\z( u \bp u + u \bp r + r \bp u  \y) + \bp \epsilon\z( u \bp u + r \p u + u \p r \y) \Big\} \\
        & + \frac{1}{\pi \aaa'} \int d^2 z~ \epsilon \z\{-\partial u \bar{\partial} u-\partial u\bar{\partial}r-\partial r\bar{\partial}u+r^2\partial\theta\bar{\partial}\theta\y\}.
    \end{aligned}
\end{equation}
The vertex is proportional to the action itself which is of order $\mathcal{O}(1)$. Therefore current conservation is broken by the source term. 
Ignoring the fact that the action is not invariant, we can write a charge from the currents that appear in the first line of \eqref{dDS}, 
\begin{equation}\label{HDL}
    H_{D} = \frac{1}{2\pi \aaa'} \oint d\sigma \z(u \p_t u + u \p_t r + r\p_t u \y).
\end{equation}
As we will show later in the Hamiltonian formalism, the charge indeed generates the dilatation transformation \eqref{D} via Poisson bracket, though it is not conserved.

\subsection{Charges in the Hamiltonian Formalism}
In this section, we follow the Hamiltonian formalism discussed in section 5 to derive the worldsheet Neother charges which generate the BMS$_3$ symmetry in asymptotically flat spacetime. The strategy is to first determine the variation of the momentum in the phase space by requiring that the worldsheet Hamiltonian and equations of motion are both preserved asymptotically. Then the infinitesimal charge can be written from \eqref{inficharge}, from which we can further obtain the finite version of the charges. From the action \eqref{actionflat}, we obtain the conjugate momentum
\begin{equation}
    \begin{aligned}
        &p_{u}=-\frac{1}{\alpha^\prime}(\partial_t u+\partial_t r),\\
        &p_{r}=-\frac{1}{\alpha^\prime}\partial_t u,\\
        &p_{\theta}=\frac{1}{\alpha^\prime} r^2\partial_t \theta.
    \end{aligned}\label{momentaflat}
\end{equation}
The symplectic form is 
\begin{equation}
    \omega=\frac{1}{2\pi}\oint d\sigma (\delta x^\mu \wedge \delta p_\mu ).
\end{equation}
The worldsheet Hamiltonian of the action \eqref{actionflat} is 
\begin{equation}\label{Hflat}
    H=\frac{\alpha^\prime}{2\pi}\oint d\sigma\z\{\frac{1}{2}p_r^2-p_rp_u-\frac{1}{2\alpha^{\prime 2}}(\partial_\sigma u)^2+\frac{p_\theta^2}{2r^2}-\frac{1}{\alpha^{\prime 2}}\partial_\sigma u\partial_\sigma r+\frac{r^2}{2\alpha^{\prime 2}}(\partial_\sigma \theta)^2\y\}.
\end{equation}
In addition, the worldsheet momentum 
\begin{equation}
    P=\frac{1}{2\pi}\oint d\sigma\z\{p_\mu \p_\sigma x^\mu\y\}\label{Pflat}
\end{equation}
is also a conserved quantity. 
Given the BMS$_3$ variation of the coordinates $\delta x^\mu =\xi^\mu$ expressed as  \eqref{bms3}, we can determine the variation of momentum $\xi^{p_\mu}$ by requiring the Jacobi identity 
\eqref{Jacobi} to be satisfied, and meanwhile the worldsheet Hamiltonian \eqref{Hflat} and momentum \eqref{Pflat} are both preserved asymptotically. More explicitly, we require the following falloff 
\begin{equation}\label{HPfalloff}
 \delta_{\xi}H\sim \mathcal{O}(r^{-2}),\quad 
  \delta_{\xi}P\sim \mathcal{O}(r^{-2}),
\end{equation} and 
\begin{equation}\label{Jaccobiflat}\{H,\xi^{I}\}-\{H_f,\{H,q^I\}\}=\z\{ \begin{array}{lll}
    \mathcal{O}(r^{-2})  & \quad  \text{for} \quad q^I=u,p_u,p_\theta  \\
    \mathcal{O}(r^{-3})  & \quad  \text{for} \quad q^I= \theta, p_r\\
    \mathcal{O}(r^{-1}) &\quad \text{for} \quad q^I=r.
\end{array} \y. \end{equation}
In the following, we will discuss the supertranslations and superrotations separately. 

\subsubsection*{Supertranslations}
Let us first consider  supertranslations in the phase space generated by 
\begin{equation}\delta u=\xi_F^u=F,\quad \delta \theta =\xi_F^\theta=-{F'\over r},\quad \delta r=\xi_F^r=F'',\quad  \delta p_\mu =\xi_F^{p_\mu}, \quad \mu=u,\theta,r,\label{supertrans}
\end{equation}
where the variation of the momentum $\xi_F^\mu$ should be determined by solving the conditions \eqref{HPfalloff} and
\eqref{Jaccobiflat}. In order to do so, we need to have an estimate of the falloff at each order. 
Using the observation of the on-shell falloff \eqref{falloff2}, we learn that for classical solutions the momentum grows at large $r$ as  
\begin{equation} \label{pfalloff}
    p_u\sim \mathcal{O}(1),\quad p_r\sim \mathcal{O}(r^{-1}),\quad p_{\theta}\sim \mathcal{O}(r).
\end{equation}
It is natural to assume the variation of momentum in the phase space $\xi_F^{p_\mu}$ grows no faster than the momentum itself ${p_\mu}$.
This allows us to find the following solution of the boundary conditions \eqref{HPfalloff} and
\eqref{Jaccobiflat}
\begin{equation}
    \begin{aligned}\label{xipT}
        & \xi^{p_u}_F=\mathcal{O}(r^{-1}),\\ 
        & \xi^{p_r}_F=-\frac{F'p_\theta}{r^2}+\mathcal{O}(r^{-2}),\\
        &\xi^{p_\theta}_F=\frac{F''p_\theta}{r}-F'p_u+\mathcal{O}(r^{-1}).
    \end{aligned}
\end{equation}
 Note that \eqref{xipT} is just the variation of the momenta if we use the on-shell relation \eqref{momentaflat} and the variation of coordinates \eqref{supertrans}, up to this order. 
Then the infinitesimal charge generating supertranslations, using the rule \eqref{inficharge}, can be written as 
\begin{equation}
    \delta H_F=-\frac{1}{2\pi}\oint d\sigma\delta\z\{Fp_u-\frac{F'p_\theta}{r}\y\}+\mathcal{O}(r^{-1}).
\end{equation}
Terms of order $\mathcal{O}(r^{-1})$ are not integrable but nevertheless can be neglected at the large $r$ limit. Thus we obtain the finite charge for supertranslation
\begin{equation}\label{HF}
    H_F=-\frac{1}{2\pi}\oint d\sigma\z\{Fp_u-\frac{F'p_\theta}{r}\y\},
\end{equation}
which agrees with the supertranslation charge  \eqref{HFL} obtained in the Lagrangian formalism.
Using Poisson brackets at larger $r$, we can also show that the stress tensor is preserved asymptotically
\begin{equation}
    \{H_F, T_{ws}\}\overset{r\rightarrow +\infty}{\approx}0,\quad \{H_F, \bar{T}_{ws}\}\overset{r\rightarrow +\infty}{\approx}0. 
\end{equation}

\subsubsection*{Superrotations }
Similarly, by solving \eqref{HPfalloff} and \eqref{Jaccobiflat}  for the superrotation
\begin{equation}
    \xi^u_Y=uY',\quad \xi^r_Y=(u Y'''-rY'),\quad \xi^\theta_Y =Y-\frac{u}{r}Y'',
\end{equation}
  we find the momentum variation
\begin{equation}
    \begin{aligned}\label{dxiY}
    & \xi_Y^{p_u}=-Y'p_u+Y''\frac{p_\theta}{r}+\mathcal{O}(r^{-1}),\\
    &\xi^{p_r}_Y=Y'p_r-\frac{uY''p_\theta}{r^2}+\mathcal{O}(r^{-2}),\\
    &\xi^{p_\theta}_Y=-Y'p_\theta+rY''p_r+\frac{uY'''p_\theta}{r}-uY''p_u+\mathcal{O}(r^{-1}),
    \end{aligned}
\end{equation}
which again agrees with directly varying \eqref{momentaflat} up to this order.
The infinitesimal charges 
\eqref{inficharge} are also integrable at the leading order, and we obtain the charges 
\begin{equation}
    H_Y=-\frac{1}{2\pi}\oint d\sigma\z\{\z(Y-\frac{uY''}{r}\y)p_\theta-rY'p_r+uY'p_u\y\}\label{srchargeH},
\end{equation}
which again agrees with the superrotation charge \eqref{HYL} obtained in the Lagrangian formalism.
At large $r$ we have
\begin{equation}
    \{H_Y, T_{ws}\}\overset{r\rightarrow +\infty}{\approx}0,\quad \{H_Y, \bar{T}_{ws}\}\overset{r\rightarrow +\infty}{\approx}0, 
\end{equation}
so that the worldsheet stress tensor is preserved asymptotically. 
\subsubsection*{Dilatation}
For the dilatation transformation \eqref{D}, there is no solution to the conditions \eqref{HPfalloff}. This is not too surprising as we have seen that the dilatation transformation does not preserve the action in the Lagrangian formalism, and therefore there is no reason to expect it to preserve the worldsheet Hamiltonian. Instead of trying to loosen the conditions \eqref{HPfalloff} and \eqref{Jaccobiflat}, we now examine directly the charge \eqref{HDL}, which we reproduce here in terms of canonical momenta, 
\begin{equation}\label{HDH}
    H_D=-\frac{1}{2\pi}\oint d\sigma\{rp_r+up_u\}.
\end{equation}
Using the Poisson bracket, we can check that $H_D$ indeed generates the transformation \eqref{D} on the coordinates,  
\begin{equation}\begin{aligned}
    \{H_D, x^\mu\}&=\xi^\mu_D.
    \end{aligned}
\end{equation}
Acting on the momenta, we get
\begin{equation}\label{xiDp}
   \xi_D^{p_u}\equiv \{H_D, p_u\}=-p_u,\quad \xi_D^{p_r}\equiv\{H_D, p_r\}=-p_r,\quad \xi_D^{p_\theta}\equiv\{H_D, p_\theta\}=0.\end{equation}
One can check that the symplectic structure and the Jacobi identity are still preserved by $H_D$ to all orders,
\begin{equation}
    \{H,\xi^I_D\}+\{H_D,\{H,q^I\}\}+\{q^I,\{H_D,H\}\}=0.
\end{equation}
On the other hand, we note that the variation generated by $H_D$ via Poisson bracket \eqref{xiDp} do not agree with direct variations of the momenta \eqref{momentaflat}. This is analogous to the case of AdS$_3$.

Recall that the dilatation transformation does not preserve the action. Relatedly, it does not preserve the Hamiltonian either, 
\begin{equation}
    \{H_D,H\}=\frac{\alpha^\prime}{\pi}\oint d\sigma\z\{-\frac{p_\theta^2}{2r^2}+\frac{r^2}{2\alpha^{\prime 2}}(\partial_\sigma \theta)^2\y\}+\mathcal{O}(r^{-1}).
\end{equation}
Therefore the dilatation does not preserve the physical phase of string theory and thus is not part of the asymptotic symmetry group.

\subsection{The BMS$_3$ algebra}
Using the mode expansion $Y_m=e^{im\theta},\quad F_m=e^{im\theta}$,  
the charges $L_m\equiv H_{Y_m}$ and $M_m\equiv H_{F_m}$ form the BMS$_3$ algebra under the Poisson bracket at large $r$ region,  
\begin{equation}
    \begin{aligned}
        &\{M_m,\, M_n\}=0,\\
        &\{L_m,\, M_n\}=-i(m-n) M_{m+n},\\
        &\{L_m,\, L_n\}=-i(m-n) L_{m+n}.
    \end{aligned}\label{bmsalg}
\end{equation}
The above BMS$_3$ algebra does not contain a central term, unlike the case of AdS$_3$. The reason is that so far we have not included any winding strings which is essential in the discussion of the central term. We leave the interesting question of finding the central charges for future study.

As mentioned before, the dilatation \eqref{D} does not belong to the asymptotic symmetry group as it does not preserve the action or the Hamiltonian. Nevertheless, we can still compute the algebra between the BMS$_3$ generators and the generator \eqref{HDH} that generates the dilation. We find that the algebra is still closed and the additional brackets are given by
\begin{equation}
{\{H_D,\, M_m\}=-M_m,\quad \{H_D,\, L_m\}=0}.\end{equation}

To summarize, in this section, we obtain the BMS$_3$ generators \eqref{bms3} from worldsheet string theory in three-dimensional flat space.
Unlike the case of AdS$_3$, variations of momenta \eqref{xipT} and \eqref{dxiY} in the Hamiltonian formalism agree with direct variations of the momenta. The Neother charges \eqref{HFL} and \eqref{HYL} agree with \eqref{HF} and \eqref{srchargeH} in the two formalisms, and they form BMS$_3$ algebra without central charges \eqref{bmsalg}. 
In addition, we also find an additional dilatation generator \eqref{D} from the asymptotic equation of motion, generated by the charge \eqref{HDH} on the phase space. The dilatation does not leave the action or Hamiltonian invariant and hence is not an asymptotic symmetry.

\section*{Acknowledgments}
We would like to thank Luis Apolo, Pengxiang Hao,  Wenxin Lai, and Xianjin Xie for useful discussions. The work is supported by the national key research and development program of China No. 2020YFA0713000.

\bibliographystyle{JHEP}      
\bibliography{ref}

\end{document}